\documentclass[prl,twocolumn,showpacs,superscriptaddress]{revtex4}

\usepackage[dvips]{graphicx}
\usepackage{amsmath}

\begin{document}

\title{Weak Ferromagnetism and Glassy State in $\kappa$-(BEDT-TTF)$_2$Hg(SCN)$_2$Br}

\author{M.~Hemmida}
\affiliation{1.~Physikalisches Institut, Universit{\"a}t Stuttgart, Pfaffenwaldring 57, 70550 Stuttgart, Germany}

\author{H.-A.~Krug~von~Nidda}
\affiliation{Experimental Physics V, Center for Electronic Correlations and Magnetism, Universit{\"a}t Augsburg, 86135 Augsburg, Germany}

\author{B.~Miksch}
\affiliation{1.~Physikalisches Institut, Universit{\"a}t Stuttgart, Pfaffenwaldring 57, 70550 Stuttgart, Germany}

\author{L. L.~Samoilenko}
\affiliation{1.~Physikalisches Institut, Universit{\"a}t Stuttgart, Pfaffenwaldring 57, 70550 Stuttgart, Germany}

\author{A.~Pustogow}
\affiliation{1.~Physikalisches Institut, Universit{\"a}t Stuttgart, Pfaffenwaldring 57, 70550 Stuttgart, Germany}

\author{S.~Widmann}
\affiliation{Experimental Physics V, Center for Electronic Correlations and Magnetism, Universit{\"a}t Augsburg, 86135 Augsburg, Germany}

\author{A.~Henderson}
\affiliation{National High Magnetic Field Laboratory, Condensed Matter Science, Florida State University, Tallahassee, Florida 32310, U.S.A.}

\author{T.~Siegrist}
\affiliation{National High Magnetic Field Laboratory, Condensed Matter Science, Florida State University, Tallahassee, Florida 32310, U.S.A.}

\author{J. A.~Schlueter}
\affiliation{Division of Materials Research, National Science Foundation, Alexandria, VA 22314, and \\
Material Science Division, Argonne National Laboratory, Argonne, Illinois 60439-4831, U.S.A.}

\author{A.~Loidl}
\affiliation{Experimental Physics V, Center for Electronic Correlations and Magnetism, Universit{\"a}t Augsburg, 86135 Augsburg, Germany}

\author{M. Dressel}
\affiliation{1.~Physikalisches Institut, Universit{\"a}t Stuttgart, Pfaffenwaldring 57, 70550 Stuttgart, Germany}

\date{\today}

\begin{abstract}
 Since the first observation of weak ferromagnetism in the charge-transfer salt $\kappa$-(BEDT-TTF)$_2$\-Cu[N(CN)$_2$]Cl [U. Welp \textit{et al.}, Phys. Rev. Lett. {\bf 69}, 840 (1992)], no further evidence of a ferromagnetic state in this class of organic materials has been reported. Here static and dynamic spin susceptibility measurements on $\kappa$-(BEDT-TTF)$_2$Hg(SCN)$_2$Br exhibit weak ferromagnetism below $20$~K on the geometrically frustrated background. Our experimental results suggest that frustrated spins in the molecular dimers suppress long-range antiferromagnetic order, forming a spin-glass type ground state of the triangular lattice in the insulating phase which locally contains ferromagnetic polarons. Moreover, specific heat data reveal an excess peak located around $5$~K indicating the glassy nature of the electrons as well.

\end{abstract}

\pacs{33.35.+r, 75.40.Gb, 76.30.-v, 75.50.Gg}

\maketitle


The search for ferromagnetism (FM) in low-dimensional organic charge-transfer salts still attracts attention in physics and chemistry. However, pure ferromagnetic organic $\pi$-electron systems containing only $s$- and $p$-valence electrons remain rare and their synthesis is a challenging problem \cite{Kahn1993}, while in inorganic materials FM usually arises from transition metals or transition-metal ions ($3d, 4f$), which in case of direct exchange may fulfill the Stoner criterion or in case of indirect exchange are subject to Goodenough-Kanamori-Anderson rules or Zener double exchange \cite{Stoner1930}.

\begin{figure}
\centering
\includegraphics[width=63mm]{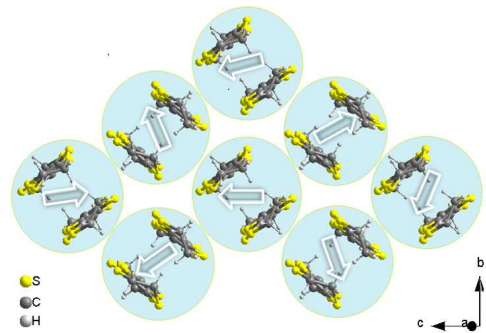}
\caption{(Color online) Top view on the conductive \textit{bc} planes of the dimerized BEDT-TTF molecules of $\kappa$-(BEDT-TTF)$_2$Hg(SCN)$_2$Br; each dimer hosts one spin. The dimer pattern can be modeled by an almost isosceles triangular lattice that is characterized by frustration effects. An artistic view of the spin arrangement illustrates that below the metal-insulator transition at $T_{\rm MI}\approx 90$~K, spin frustration becomes dominant and suppresses the magnetic order.}
\label{Structure}
\end{figure}

The organic radical salts $\kappa$-(BEDT-TTF)$_2$\textsl{X}, where BEDT-TTF is the abbreviation of bis-(ethylene\-di\-thio)\-tetra\-thia\-fulvalene, consist of alternating layers of the electron donor BEDT-TTF and electron acceptor $X$. In the $\kappa$-phase crystal structure the BEDT-TTF molecules stack in pairs as depicted in Fig.~\ref{Structure}. Here the (BEDT-TTF)$_2$ dimers are arranged in a two-dimensional structure rather than in chains. Within the BEDT-TTF layers, the molecular dimers are close to each other, allowing substantial overlap of the molecular orbitals. Since one electron is transferred from each (BEDT-TTF)$_2$ dimer to the anion, the conduction band is half-filled. For weak electronic correlations, this implies that these organic compounds are metallic enabling nearly isotropic electron motion within the layer; perpendicular to the plane the resistivity is larger by more than one order of magnitude \cite{Ishiguro1998,Mori2016}.

The BEDT-TTF-based salts can be easily tuned by hydrostatic and uniaxial pressure, deuteration, or chemical substitution such that a wide range of electronic phases is obtained including paramagnetic, antiferromagnetic (AFM), spin-liquid, and superconducting ground states \cite{Powell2011}. The best studied examples are the $\kappa$-(BEDT-TTF)$_2X$ compounds, where $X$ denotes the anion such as Cu[N(CN)$_2$]Cl$^-$, Cu[N(CN)$_2$]Br$^-$ or Cu$_2$(CN)$_3$$^-$ \cite{Kanoda1997}. For these compounds the onsite Coulomb repulsion $U$ is comparable to the bandwidth $W$, placing them close to the Mott metal-insulator transition \cite{Pustogow2018}.

Here we want to draw the attention to the family of $\kappa$-type BEDT-TTF salts with mercury-based anions \cite{Lyubovskii1996} where the ratio of Coulomb interaction and bandwidth is significantly smaller; correspondingly the inter-dimer coupling is stronger than in the Cu-based compounds \cite{Drichko2014,Ivek2017}. Electrical resistivity measurements evidence that $\kappa$-(BEDT-TTF)$_2$Hg(SCN)$_2$Br undergoes a metal-insulator transition (MIT) upon cooling below the transition $T_{\rm MI}$; the origin of the MIT, however, is still under debate \cite{Ivek2017,Loehle2017}. The isostructural sister compound $\kappa$-(BEDT-TTF)$_2$Hg(SCN)$_2$Cl enters a charge-ordered state at $T_{\rm CO}=30$~K \cite{Drichko2014}.

Comprehensive transport, dielectric, and optical investigations of the present compound $\kappa$-(BEDT-TTF)$_2$Hg(SCN)$_2$Br, however, do not find any evidence of charge order or structural changes \cite{Ivek2017}. This might resemble the Mott insulators $\kappa$-(BEDT-TTF)$_2$Cu[N(CN)$_2$]Cl, $\kappa$-(BEDT-TTF)$_2$Cu$_2$(CN)$_3$ or $\kappa$-(BEDT-TTF)$_2$Ag$_2$(CN)$_3$, where also no indication of charge order is seen \cite{Sedlmeier2012,Pinteric2016}, yet the resistivity jump in $\kappa$-(BEDT-TTF)$_2$Hg(SCN)$_2$Br is rather abrupt.

In an early electron spin resonance (ESR) characterization of hydrogenated and deuterated $\kappa$-(BEDT-TTF)$_2$\-Hg(SCN)$_2$Br reported two decades ago \cite{Yudanova1995}, a first-order phase transition was supposed in the hydrogenated compound around $100$~K and associated with localization of electrons on the (BEDT-TTF)$_2$ dimers; a transition of semiconductor-semiconductor type was suggested for the deuterated compound with possible magnetic ordering. Based on detailed magnetization and ESR investigations we were now able to unveil a weak FM state at low temperatures. There is no long-range order, instead the frustrated spins form a spin glass-type state. Even beyond that, specific heat measurements suggest the existence of the electron-glass state.

Single crystals of $\kappa$-(BEDT-TTF)$_2$Hg(SCN)$_2$Br were prepared following the synthesis method of Lyubovskaya and collaborators \cite{Konovalikhin1992,Aldoshina1993}. It is important to note that the stable divalent state of the Hg$^{2+}$ ions $(5d^{10})$ is non-magnetic and, hence, magnetism will not be hampered by any kind of valence changes like in the case of the Cu-based BEDT-TTF salts; besides a majority of Cu$^{+}$ ions $(3d^{10})$ those crystals usually contain Cu$^{2+}$ ions $(3d^{9})$, which can strongly influence the magnetic properties \cite{Padmalakha2015}.

Magnetization measurements were performed at temperatures $2 \leq T \leq 300$~K using a superconducting quantum interference device MPMS XL (Quantum Design). As the mass of the single crystals under inspection is only about $1$~mg, the magnetization data had to be corrected by subtraction of the diamagnetic background of the sample holder, measured beforehand. The diamagnetic contribution of $\kappa$-(BEDT-TTF)$_2$Hg(SCN)$_2$Br is estimated as $\chi_{\rm dia} \approx -9\times10^{-4}$~emu/mol taking into account the bond structures of the (BEDT-TTF) molecule (see Ref.~\onlinecite{Dorfman1965}).
Specific heat measurements were performed by a Physical Properties Measurement System (Quantum Design) for temperatures $1.8 \leq T \leq 300$~K and magnetic fields up to $3$~T. Beside the standard technique, a large heat-pulse method was applied in order to properly probe the first-order MIT \cite{Widmann2016}.
The ESR measurements were performed in a Bruker X-band spectrometer equipped with a continuous He-gas flow cryostat working in the temperature range down to $4.2$~K. The samples were fixed in a quartz tube by paraffin and could be rotated by a goniometer.

\begin{figure}
\centering
\includegraphics[width=65mm]{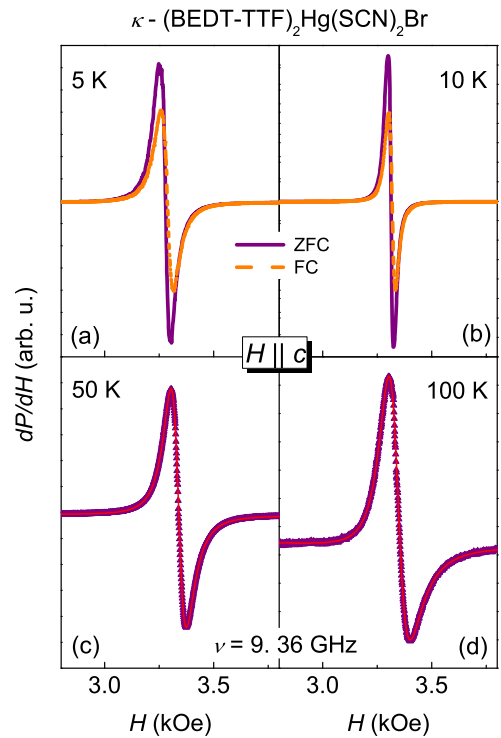}
\caption{(Color online) ESR spectra, i.e. field derivative of absorbed power, of $\kappa$-(BEDT-TTF)$_2$Hg(SCN)$_2$Br taken at X-band frequency for selected temperatures and $H\parallel c$ as indicated. Zero-field cooled (ZFC) and field cooled (FC) measurements (orange) are depicted in (a) and (b). The red solid lines, shown in (c) and (d) correspond to a fit by the field derivative of a Lorentz and Dyson line, respectively.}
\label{Spectra}
\end{figure}

In the whole temperature regime, the ESR spectra of $\kappa$-(BEDT-TTF)$_2$Hg(SCN)$_2$Br are well described by the field derivative of a single Lorentz/Dyson line [Fig.~\ref{Spectra}]. The signal results from magnetic dipole transitions between the Zeeman levels of the conduction-electron spins. Above $T_{\rm MI} \approx 90$~K the system is metallic and especially for the case that the microwave field is applied perpendicular to the conductive planes, the Lorentz line transforms to an asymmetrical shape (Dysonian) [Fig.~\ref{Spectra}(d)]. This is due to the skin effect, which appears in conductive compounds because of the interaction between the applied microwave field and mobile charge carriers. For the microwave field applied perpendicular to the conductive \textit{bc} planes, shielding currents are induced in these conductive planes that drive electric and magnetic microwave components out of phase. This yields an admixture of dispersion to the absorption depending on the ratio of skin depth and sample size \cite{Barnes1981}. In the insulating phase the signal has a symmetrical shape [Fig.~\ref{Spectra}(c)]. Below $T_{\rm g} \approx 40$~K comparison of zero-field cooled (ZFC) and field cooled (FC) measurements reveals a slight resonance shift ($0.5$~Oe at $30$~K, $10$~Oe at $5$~K), line broadening ($0.3$~Oe at $30$~K, $8$~Oe at $5$~K) and decrease of the amplitude for the FC case [Fig.~\ref{Spectra}(a,b)]. This is a well-known aspect of a spin glass behavior (see e.g Ref.~\onlinecite{Upadhyay2003}). Thus $T_{\rm g}$ can be identified with the glass-transition temperature. Note that the FC effect is fully developed for cooling fields larger than $1$~kOe. One may argue that a pure FC/ZFC effect is not unique to spin glasses, but could also result from domain reorientation. However, domain reorientation implies long-range spin order which has been discarded at the MIT previously (see Ref.~\onlinecite{Ivek2017}). No further transition is observed in the specific heat down to 2 K as will be shown below.

\begin{figure}
\centering
\includegraphics[width=69mm]{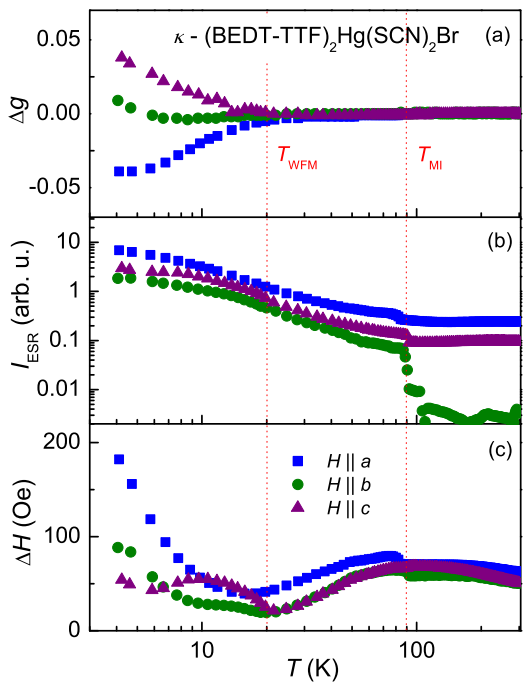}
\caption{(Color online) Temperature dependence of ESR parameters (FC) of $\kappa$-(BEDT-TTF)$_2$Hg(SCN)$_2$Br for the magnetic field applied along the crytallographic principal axes: (a)~$g$-shift $\Delta g = g-g(300$~K), (b)~double integrated intensity $I_{\rm ESR}$, and (c)~linewidth $\Delta H$ determined at X-band frequency. Red dotted lines indicate $T_{\rm MI}$ at $90$~K and a weak ferromagnetic transition at $T_{\rm WFM} \approx 20$~K.} \label{ESRparameters}
\end{figure}

The ESR parameters determined from the fits of the spectra measured in all three crystal directions are displayed in Fig.~\ref{ESRparameters} as a function of the temperature in the upper panel (a) the $g$-shift $\Delta g$, followed by the signal intensity $I_{\rm ESR}$, and the linewidth $\Delta H$ in the lowest frame (c). The most prominent effect of the MIT is visible in the ESR intensity. Resembling the spin susceptibility, in the metallic regime it is Pauli-like, i.e.\ approximately constant; a step indicates the MIT around $90$~K. In the insulating regime a pronounced monotonous increase is observed with the tendency for saturation to lowest temperatures. Note that the step is largest for $H\parallel b$ where the microwave field is oscillating perpendicular to the conductive layers, giving rise to the strongest shielding in the metallic regime. The linewidth amounts to about $50-60$~Oe at room temperature, dependent on the orientation of the sample in the magnetic field, and increases only slightly upon decreasing temperature. As seen in Fig.~\ref{ESRparameters}(c), on approaching $T_{\rm MI}$, $\Delta H$ increases abruptly by about $10$~Oe and then decreases significantly below $50$~K. Finally, for $T < 20$~K, one clearly recognizes a strong broadening of $\Delta H$. The $g$ values are close to $2$ at $300$~K ($g_{\rm a} = 2.011$, $g_{\rm b} = 2.004$, $g_{\rm c} = 2.004$) and remain nearly unchanged on passing $T_{\rm MI}$, but develop a strong anisotropy below $20$~K, as plotted in Fig.~\ref{ESRparameters}(a).

\begin{figure}
\centering
\includegraphics[width=69mm]{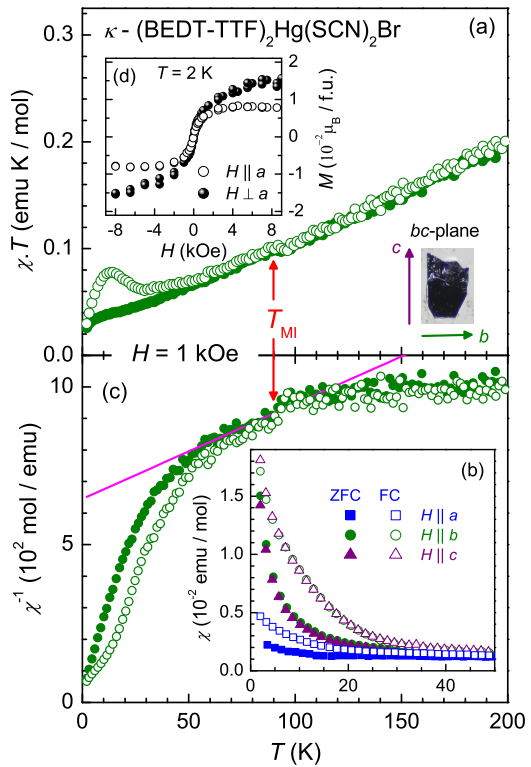}
\caption{(Color online) (a)~Temperature dependence of the susceptibility $\chi$ of $\kappa$-(BEDT-TTF)$_2$Hg(SCN)$_2$Br in representation $\chi \cdot T$. (b)~$\chi(T)$ for the magnetic field $H=1$~kOe applied along the three principal crystallographic axes under zero-field cooled (ZFC, solid symbols) and field cooled (FC, open symbols) conditions. The inset depicts the single crystal under investigation. Note that the \textit{bc}-plane is the conductive plane (c)~Inverse susceptibility as a function of temperature. Straight lines indicate Curie-Weiss laws as discussed in the text. The inset (d) shows the field dependence of the magnetization at $T = 2$~K.}
\label{SpinSusceptibility}
\end{figure}

For a quantitative analysis we consider complementary measurements of the static susceptibility $\chi$. In Fig.~\ref{SpinSusceptibility}(a), we chose the $\chi \cdot T$ plot in order to elucidate the different temperature regimes: The strictly linear increase above $T_{\rm MI}$ characterizes the Pauli-like paramagnetism. Just below $T_{\rm MI}$ the descent of the curve is typical for dominant AFM exchange interaction. On further decreasing temperature the positive curvature, which for FC conditions is most pronounced and results even in a local maximum of $\chi \cdot T$ below approximately $20$~K, indicates competing FM exchange as described e.g in Ref.~\onlinecite{Mori2016}. Turning to Fig.~\ref{SpinSusceptibility}(c), in the temperature range $50 < T < 90$~K the system follows a strongly AFM Curie-Weiss law $\chi = C/(T-\Theta_{\rm CW})$ (magenta line) with a Curie-Weiss temperature $\Theta_{\rm CW} = -215(70)$~K and a Curie constant $C=0.33$~emu/(mol\,K) corresponding to $0.9(0.25)$ electron spins per (BEDT-TTF)$_2$ dimer; hence within the experimental uncertainty all electron spins contribute to the susceptibility. As seen in Fig.~\ref{SpinSusceptibility}(b) and (c), ZFC and FC data become distinct when the temperature is reduced below $50$~K with a strongly increasing slope. The FC data follow an S-shape curve, which finally joins the ZFC data at $2$~K. As shown in the inset of Fig.~\ref{SpinSusceptibility}(d), the field-dependent magnetization reveals a soft FM loop with a saturation at about $15\times10^{-3}~\mu_{\rm B}$/formula unit. This value is one order of magnitude larger than that found in $\kappa$-(BEDT-TTF)$_2$Cu[N(CN)$_2$]Cl ($8\times10^{-4}$$\mu_{\rm B}$/formula unit) \cite{Welp1992}. Moreover the susceptibility evolves a significant easy-plane anisotropy to low temperatures which is also visible in the magnetization data [see Fig.~\ref{SpinSusceptibility}(b,d)].

\begin{figure}
\centering
\includegraphics[width=69mm]{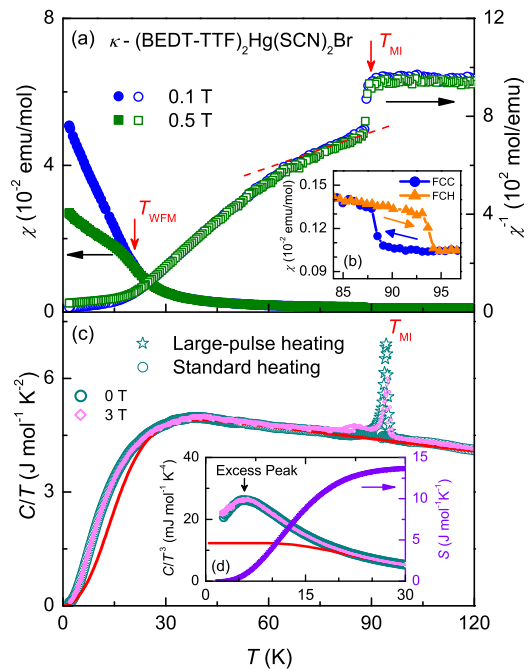}
\caption{(Color online) (a) Temperature-dependence of the magnetic susceptibility of an assembly of crystals (left ordinate) and its inverse (right ordinate) measured at two different magnetic fields. (b) Hysteresis of the temperature dependent susceptibility on cooling and heating through the MIT. (c) Temperature dependence of specific heat in representation of $C/T$. The red solid line corresponds to Debye-Einstein fit (see text). (d) $C/T^3$ vs $T$ plot reveals an excess-heat capacity or excess peak. Both MIT and excess peak are robust at $3$~T.The entropy $S$ is obtained as $13.7$~J/mol.K}
\label{Boson}
\end{figure}

FC susceptibility measurements in different fields for an assembly of several crystals support and complete the findings in the single crystal (Fig.~\ref{Boson}(a)). The sharp step at $90$~K marks the MIT and its first-order character revealed by the corresponding hysteresis on cooling and heating through the transition (Fig.~\ref{Boson}(b)). Below the MIT the slope  of the inverse susceptibility (indicated as a dashed line) agrees with the Curie-Weiss constant found in the single crystal independent of the applied magnetic field. On further cooling, below $20$~K a field dependence shows up characterizing the weak ferromagnetism present in the glassy regime.

Our comprehensive experimental results on $\kappa$-(BEDT-TTF)$_2$Hg(SCN)$_2$Br suggest, first, that the MIT is not accompanied by magnetic long-range order, and second, around $20$~K a weak ferromagnetic phase develops. In the following we will discuss these two points: the Curie-Weiss law observed in the temperature range $50 < T < 90$~K, with all electrons contributing, provides clear evidence that no long-range order exists below $T_{\rm MI}$. This behavior is characteristic for a paramagnetic phase of localized spins. The large negative $\Theta_{\rm CW}$ indicates strong AFM exchange interactions between the spins. While in the case of the Cu- and Ag-based $\kappa$-BEDT-TTF salts, $J/k_{\rm B}=200-300$~K is reported \cite{Shimizu2003}, for $\kappa$-(BEDT-TTF)$_2$Hg(SCN)$_2$Br we estimate $J/k_{\rm B}\approx 70$~K from mean-field theory \cite{Smart1966}. This value is in a good agreement with that estimated for a dipole solid using a tight-binding approximation ($J/k_{\rm B}=80$~K) \cite{Hassan2017}.

We can definitely rule out any pairing of the spins, as they would not contribute to the susceptibility. Moreover, it is important to note that the $g$-values do not change at $T_{\rm MI} \approx 90$~K, i.e.\ the average position of the electron on the BEDT-TTF molecule does not depend on its mobility. This finding strongly supports the absence of charge order deduced from vibrational spectroscopy \cite{Ivek2017}. When the electrons become localized, they remain randomly distributed on the (BEDT-TTF)$_2$ dimers, but do not occupy a certain molecular site.
The triangular structure of the (BEDT-TTF)$_2$ layers causes strong geometrical frustration of the AFM exchange interaction. The deviation of both static susceptibility and ESR intensity from the Curie-Weiss law below $50$~K probably arises from critical magnetic fluctuations and short-range order. In particular for $T < 20$~K a significant increase of the magnetic susceptibility is observed. The comparison of FC and ZFC susceptibilities and ESR spectra as well as the field dependence of the magnetization at low $T$ indicates a weak ferromagnetic polarization. The observed easy-plane anisotropy is typical for thin ferromagnetic layers due to the demagnetization. In $\kappa$-(BEDT-TTF)$_2$Hg(SCN)$_2$Br frustration -- due to the triangular lattice -- and disorder -- due to the absence of charge order -- are both present; hence, a spin-glass-like ground state can be expected as sketched in Fig.~\ref{Structure}.

Focusing on the glassy state, the specific heat data provide further important hints [Fig.~\ref{Boson}(c,d)]. The phonon contribution was estimated using one Debye contribution with a characteristic Debye temperature of $\Theta_{\rm D} = 122$~K and two Einstein modes corresponding to temperatures of $\Theta_{\rm E1} = 300$~K and $\Theta_{\rm E2} = 600$~K with nearly equal weights. Notably, the Einstein temperatures nicely correspond to two prominent phonon modes at $200$~cm$^{-1}$ and $430$~cm$^{-1}$ as given in Ref.~\onlinecite{Ivek2017}. The first-order transition peak around $T_{\rm MI}$ could not be clearly detected by the standard technique due to the latent heat. Therefore we added data of the large-pulse method. Most important below $30$~K, a significant non-Debye behavior points toward low-energy vibrational states, which are not accounted for by the Debye-Einstein model [Fig.~\ref{Boson}(d)]. This characteristic excess-heat capacity or excess peak corresponds to the Boson peak observed in scattering experiments and is a characteristic and universal feature of disordered matter \cite{Phillips1981}. Both excess peak and MIT are found to be robust against strong external magnetic fields. This means that the MIT and glassy freezing are basically related to the electron charge dynamics. Indeed the entropy contained in the excess peak $S = 13.7$~J/mol.K, is significantly larger than the pure magnetic entropy of the spin-1/2 system ($S = R \ln 2$). It indicates that the glassy state is mainly engendered by the disorder of charge degree of freedom, accompanied by the disordered spin degree of freedom resulting in a charge-spin entanglement.

Coulomb repulsion favors a uniform electronic density, while disorder drives local density fluctuations. When these two effects are comparable in magnitude, one expects many different low-energy electronic configurations, i.e.\ many metastable states in the insulating side of the MIT \cite{Dobrosavljevic2012}. Similar to a spin glass in frustrated spin systems with disorder, the emergence of such metastable states leads to an electron-glass state, characterized by slow relaxation dynamics and manifested in the specific-heat data as an excess peak. Indeed in the geometrically frustrated compound $\theta$-(BEDT-TTF)$_2$RbZn(SCN)$_4$ \cite{Kagawa2013}, the resistivity at low temperatures behaves in a similar way as in $\kappa$-(BEDT-TTF)$_2$Hg(SCN)$_2$Br (Ref.~\onlinecite{Ivek2017}). Such behavior is attributed to the formation of a charge-cluster glass, i.e\ electrons struggle by frustration. Here we recall the Raman data in Ref.~\onlinecite{Hassan2017}, which states the absence of charge order due to quantum electric dipole fluctuations within the molecular dimers. Early theoretical studies of similar systems predicted a quantum melting of charge-order due to the frustration effect \cite{Merino2005}.

In order to understand the impact of quantum melting on the glassy state below $T_{\rm MI}$, we refer to recent theoretical studies which find that quantum tunneling broadens the supercooled liquid regime in low temperature glass formers and causes a significant decrease of $T_{\rm g}$ with respect to $\Theta_{\rm D}$ ($T_{\rm g}/\Theta_{\rm D}<0.5$) \cite{Novikov2013}. The Boson peak observed by Raman spectroscopy in some hydrated biomolecules verified these theoretical predictions \cite{Lima2014}. In $\kappa$-(BEDT-TTF)$_2$Hg(SCN)$_2$Br, the value of $T_{\rm g}/\Theta_{\rm D}=40/122\approx0.33$ suggests the electron tunneling within the (BEDT-TTF)$_2$ dimers. This process melts the electron glass and leads to an emergent quantum dipole liquid state.

Finally, back to the origin of weak ferromagnetism, it is important to note that in antiferromagnetic semiconductors due to the competition between kinetic energy of conduction electrons and antiferromagnetic exchange, an electron may become localized in a magnetic microregion. This microregion may be ferromagnetic called a ferromagnetic polaron or ferron \cite{Nagaev1983}. The first experimental evidence of ferrons has been found in EuTe and EuSe \cite{Wachter1972}. Here we suggest that a similar mechanism drives weak ferromagnetism, however theoretical work is needed to clarify this issue.

To summarize, we found that on passing the MIT in $\kappa$-(BEDT-TTF)$_2$Hg(SCN)$_2$Br, the conduction electrons localize at the (BEDT-TTF)$_2$ molecular dimers, but without formation of a magnetic long-range order. Instead, geometric frustration keeps the AFM coupled spin system in a paramagnetic state, as proven by the Curie-Weiss law of the susceptibility. Taking into account that there is no charge order around $T_{\rm MI}$, disorder in the related spin-density of neighboring (BEDT-TTF)$_2$ molecular dimers locally gives rise to weak ferromagnetism for $T < 20$~K. Regarding the FC induced enhancement of the susceptibility and the excess peak in the specific-heat data at temperatures below $10$~K, we suggest that the evolution of the glassy state of both electrons and spins in the presence of strong quantum fluctuations, results in an entanglement between spin and charge degrees of freedom.

\acknowledgments
This work was supported by the Deutsche Forschungsgemeinschaft (DFG) by DR228/39-1 and DR228/52-1, as well as the Transregional Collaborative Research Center TRR 80 (Augsburg-Munich-Stuttgart). JAS acknowledges support from the Independent Research and Development program from the NSF while working at the Foundation.


\end{document}